\documentclass[12pt]{article}

\usepackage[hmargin=2cm,vmargin=1.5cm]{geometry}
\usepackage{graphicx}
\usepackage{amsmath}
\usepackage{enumerate}
\begin{document}
\title{Coherences of accelerated detectors and the local character of the Unruh effect}
\author {Charis Anastopoulos\footnote{anastop@physics.upatras.gr} \\
 {\small Department of Physics, University of Patras, 26500 Greece} \\ \\
 and \\ \\ Ntina Savvidou\footnote{ntina@imperial.ac.uk} \\
  {\small  Theoretical Physics Group, Imperial College, SW7 2BZ,
London, UK} }
\maketitle
\begin{abstract}
We study   the locality of the acceleration temperature in the Unruh effect. To this end, we develop a new formalism for the modeling of macroscopic irreversible detectors. In particular, the formalism allows for the derivation of the higher-order coherence functions, analogous to the ones employed in quantum optics, that encode temporal fluctuations and correlations in particle detection. We derive a causal and approximately local-in-time expression for an Unruh-Dewitt detector moving in a general path in Minkowski spacetime. Moreover, we derive the second-order coherence function for uniformly accelerated Unruh-Dewitt detectors. We find that the fluctuations in detection time for a single Unruh-Dewitt detector  are thermal. However, the correlations  in detection-time between two Unruh-Dewitt detectors with the same acceleration but separated by a finite distance are {\em not} thermal. This result  suggests that the Unruh effect is fundamentally local, in the sense that the notion of acceleration temperature applies only to the properties of local field observables.

\end{abstract}

\section{Introduction}
A fundamental property of quantum field theories on Minkowski spacetime is that the number of particles in a given quantum field state is the same for all inertial observers. However, this equivalence does not hold for non-inertial observers, because such observers define particles with respect to different normal modes of the field \cite{Full, Unruh2}. This fact is aptly demonstrated by the Unruh effect: for an observer moving with uniform proper acceleration $a$, the usual quantum field vacuum appears as a heat bath at temperature $T = \frac{a}{2\pi}$. The relation between acceleration and temperature, characterizing the Unruh effect, has strong analogies to the particle emission from black holes \cite{Hawk} and cosmological horizons \cite{GibHawk} and for this reason, it constitutes a fundamental ingredient of theories advocating a  thermodynamic origin of gravity \cite{thermgrav}.

 In quantum field theory, the Unruh effect follows from the fact that the restriction of the Minkowski vacuum in one Rindler wedge is shown to satisfy the Kubo-Martin-Schwinger (KMS) condition for thermal states \cite{BW}. This derivation employs the idealized notion of eternally accelerated observers and it depends on global properties, such as the existence of a Rindler horizon \cite{Full2}. For this reason, it is important to confirm the physical relevance of acceleration temperature in models that involve only local physics \cite{Unruh, UW84, BL, AM, matsas, review}. The most commonly employed model is the Unruh-Dewitt detector, introduced in Ref. \cite{Dewitt}, which consists of a point-like monopole interacting with a quantum field and moving with uniform acceleration in Minkowski spacetime.

In this paper, we approach the Unruh effect by
studying the response of accelerated {\em macroscopic} detectors, analogous to ones that have been long employed in quantum optics \cite{Glau}. The use of macroscopic detector models allows for the consideration of more general observables that correspond to the temporal fluctuations and correlations for the recorded particles. In particular, we
construct the second-order coherence function for accelerated detectors that encodes the  fluctuations of their recorded intensity \cite{WM}.  We find that the intensity fluctuations recorded by a single accelerated detector are thermal, in the sense that the corresponding second-order coherence function  coincides with that of a static detector in a thermal bath. However, the intensity correlations of a  pair of Unruh-DeWitt detectors, moving with the same acceleration along different paths are {\em not}  thermal: they differ significantly from the intensity correlations recorded by a pair of static detectors in a thermal bath.

The results above emphasize the {\em local} character of acceleration temperature, in the following sense. The intensity fluctuations of a single detector correspond to local field observables, while the correlation between spatially separated detectors correspond to non-local field observables. Therefore, we conclude that the equivalence between acceleration and temperature persists for local observables, even when higher-order correlations are taken into account, but it is lost when one considers non-local physical observables.

The consideration of macroscopic detectors is crucial for the derivation of the correlations and fluctuations of the recorded particles, and it is the key point of departure of this paper from past works on the topic, where detectors are treated as microscopic systems.
Microscopic systems are meaningful {\em probes} of a quantum field, and as such they suffice to ascertain the existence of acceleration temperature,  but they do not  conform to the conventional concept of a detector \cite{Pad0}. An actual detection event involves an irreversible amplification of microscopic processes, a definite macroscopic record and it is well localized in time.

Using only microscopic probes, it is impossible to associate a definite instant of time (or time interval) with a detection event. An elementary example that demonstrates this point is the following. Let us consider a single (isolated) atom interacting with a quantum field through a dipole coupling. In absence of a measurement, i.e., an interaction with a macroscopic measuring device that leaves a definite macroscopic record, we cannot tell when during its history, the atom absorbed a photon. A definite   record   requires that (at least part of the) detecting system be macroscopic. One may consider for example a  detector  consisting  of an array of atoms, together with a macroscopic photo-detector.   When an atom absorbs a photon, it becomes excited. However, the excited state has a finite lifetime $\Gamma^{-1}$; the atom then re-emits a photon which leaves a macroscopic record at the photo-detector. Thus, the moment of detection can be localized in time with an uncertainty of order $\Gamma^{-1}$. In absence of the macroscopic photo-detector, the instant of detection cannot be specified.

In existing treatments, the response of a microscopic probe coupled to a scalar field is determined from the solution of Schr\"odinger's equation for the total system.
 To first order in perturbation theory, the transition amplitude for a uniformly accelerated Unruh-Dewitt detector corresponds to a Planckian spectrum \cite{Dewitt}. Higher-order effects, such as dissipation and noise, destroy the thermal behavior for proper times much larger than the decay time scale  of the detector \cite{HuLin}. This problem does not arise if one assumes that the detector-field interaction is  switched on only  for a finite time interval \cite{SvSv, Hig93, SPad96} of duration $T$ much smaller than the decay time scale. In this case, the  transition amplitude refers to the state of the detector after the interaction has been switched-off: it does not define the detection probability {\em during} the period of interaction. To first order in perturbation theory, the transition amplitude is proportional to the time $T$, hence, it is possible to infer  an expression for the detector's excitation rate at intermediate times. With a suitable regularization, one  obtains a causal expression for the excitation rate of a detector moving along general trajectories in Minkowski spacetime \cite{Schlicht, Lang, LouSa}.

 However, a purely unitary evolution of a microscopic system cannot account for the irreversible macroscopic amplification corresponding to a detection event and the creation  of an associated record. This implies, in particular, that correlations between detection events cannot be determined.  For example, the joint probability that a particle is detected at time $t_1$ and another particle is detected at time $t_2$ cannot be constructed. In order to construct such joint probabilities, it is necessary to incorporate the information obtained by the first detection into the quantum state--- as for example, when one considers sequential measurements \cite{Dav, Omn2, Ana1}.

Multi-time correlations are observable in quantum optics \cite{WM}, where the statistical analysis of photon-detection times  provides significant information about the  quantum field state. The photon statistics may be Poissonian, sub-Poissonian, or super-Poissonian, and the photons may either bunch or anti-bunch. (See, Ref. \cite{WM} for a definition of these terms,  and Sec. 3.2.2 for their adaptation in the present context.) The measurement of such correlations requires macroscopic detectors with well-defined records for the instants of detection. These  properties must also be reflected in the theoretical description of the detection process.

To this end, we employ techniques from the quantum theory of measurement \cite{Dav, WZ,BLM} in order to construct a general model for macroscopic particle detectors, moving along general trajectories in Minkowski spacetime. In particular, we adapt a method that was developed in Refs. \cite{AnSav, AnSav2} for the study of  time-of-arrival and related problems in quantum theory. This method contains some ideas and techniques from the consistent/decoherent histories approach to quantum mechanics \cite{Gri, Omn, Omn2, GeHa, hartlelo}, and it is related to the Davies-Srinivas photodetection theory \cite{SD81}. The most significant advantage of this method is that it leads to the definition of detection probabilities where the time of detection $t$ is treated as a genuine {\em quantum variable}, and not as an external parameter. In particular, the detection probabilities are densities with respect to $t$ and they define  a Positive-Operator-Valued-Measure (POVM). The treatment of detection time as a quantum observable is  essential for an unambiguous and probabilistically sound definition of the  fluctuations and correlations with respect to the time of detection.

 Using this method, we obtain general expressions for the detector's response, which are causal and approximately local in time. We also derive expressions for the multi-time joint probabilities of particle detection. The correlations for a pair of uniformly accelerated detectors are thermal if the detectors are co-incident, but non-thermal if they are separated by a finite distance $d$.

The structure of the paper is the following. In Sec. 2, we develop a general formalism  that determines the intensity and the intensity fluctuations of irreversible particle detectors. In Sec. 3, we apply this formalism to detectors with a dipole coupling to a massless scalar field (Unruh-DeWitt) detectors. We derive expressions for the detection  probabilities and correlations for such detectors moving along general trajectories in Minkowski spacetime, and we then specialize to the case of constant acceleration. In Sec. 4, we summarize our results, emphasizing the local character of acceleration temperature.

\section{Probabilities for an irreversible particle detector}

In this section, we develop a general formalism that determines the joint probabilities for a sequence of quantum events, with respect to the times  occurring at different time instants $t_1, t_2, \ldots, t_n$. The novelty in this construction is that the times $t_1, t_2, \ldots, t_n$ are treated as genuine quantum observables. The probabilities are densities with respect to $(t_1, t_2, \ldots, t_n)$ and they correspond to a Positive-Operator-Valued-Measure. These probabilities are then employed in order to derive the
intensity and the intensity fluctuations of irreversible, macroscopic particle detectors.

The main requirement in this construction is the existence a coarse-graining time-scale $\sigma$, such that all temporal interferences at time-scales larger than $\sigma$ are suppressed. In the language of decoherent histories, this means that histories corresponding to a transition taking place at some time interval $[t_i, t_f]$ are decoherent, if $|t_f - t_i| >> \sigma$. This property is expected to hold for macroscopic systems, or systems that possess a macroscopic component \cite{Gri, Omn, GeHa}.

\subsection{Probability amplitudes for the time of detection}

 The first step in our construction is the derivation of a general expression for the single-time probability for a  quantum event taking place at time $t$. The event time $t$ is treated as a quantum observable, and the probabilities are densities with respect to $t$. The more general case of multi-time probabilities  associated to a sequence of quantum events is treated in Sec. 2.3.

  First, we  derive an expression for
 the probability amplitude that, given an initial state $|\psi_0\rangle$, a transition occurs at some instant in the time interval $[t_i, t_f]$ and a recorded value $\lambda$ is obtained for some observable.

Let ${\cal H}$ be the
Hilbert space of a quantum system. Since we are interested in describing measurements, ${\cal H}$ includes the degrees of freedom of a macroscopic apparatus in addition to the ones of a microscopic system.
In order to describe a quantum event, like a particle detection, we assume that
 ${\cal H}$ splits into two subspaces: ${\cal H} = {\cal
H}_+ \oplus {\cal H}_-$. The subspace ${\cal H}_+$ describes the accessible
states of the system given that a specific event is realized; the subspace ${\cal H}_-$ is the complement of ${\cal H}_+$. For example, if the quantum event under consideration is a detection of a particle by a macroscopic measuring device, ${\cal H}_+$ describes the possible states of the system provided that the detection has occured and ${\cal H}_-$ describes the possible states of the system if the detection has not occurred.
We denote  the
projection operator onto ${\cal H}_+$ as $\hat{P}$ and
the projector onto ${\cal H}_-$ as $\hat{Q} := 1  - \hat{P}$.

Once the transition has taken place, it is possible to
measure the values of various observables through their correlation to a pointer variable. We denote by $\hat{P}_\lambda$   projection operators corresponding to different values $\lambda$ of an observable that can be measured only if the quantum event under consideration has occurred. For example, when considering a particle detection, $\hat{P}_\lambda$ may correspond to possible values of a pointer variable in the apparatus that is correlated to some microscopic property of the system: position, momentum, energy, and so on.
The set of projectors $\hat{P}_\lambda$ is exclusive ($\hat{P}_{\lambda} \hat{P}_{\lambda'} = 0, $ if $\lambda \neq \lambda'$). It is also exhaustive given that the event under consideration  has occurred; i.e., $\sum_\lambda \hat{P}_\lambda = \hat{P}$.

We assume that  the system is initially (at time $t = 0$) prepared in a
state $|\psi_0 \rangle \in {\cal H}_+$, and that the dynamics is
governed by the self-adjoint Hamiltonian operator $\hat{H}$.

We proceed to derive an expression for the  amplitude $| \psi; \lambda, [t_1, t_2] \rangle$, that a transition occurs at a time interval $[t_1, t_2]$ and a recorded value $\lambda$ is obtained.
 We first consider the case that the relevant time
interval is small, i.e., we set $t_1 = t$ and $t_2 = t + \delta t$, and we  keep only terms of leading  to $\delta t$.
Since the transition takes place within the interval $[t, t + \delta
t]$, at times prior to $t$ the state lay within ${\cal H}_-$. This is taken into account by evolving the initial state $|\psi_0 \rangle$ with the
restriction of the propagator into ${\cal H}_-$, that is, with the operator
\begin{eqnarray}
\hat{S}_t =  \lim_{N \rightarrow \infty}
(\hat{Q}e^{-i\hat{H} t/N} \hat{Q})^N. \label{restricted}
\end{eqnarray}

By assumption, the transition took place some time within the time interval $[t, t+\delta t]$ and at the end  the outcome $\lambda$ is recorded. This means that in the time-interval $[t, t+\delta t]$ the amplitude transforms under the
full unitary operator for time evolution  $e^{-i \hat{H} \delta t}
\simeq 1 - i \delta t \hat{H}$. At time $t + \delta t$ the event
corresponding to $\hat{P}_{\lambda}$ is recorded, so the amplitude
is transformed by the action of $\hat{P}_{\lambda}$ (or of $\sqrt{\hat{P}_{\lambda}}$, if $\hat{P}_{\lambda}$ is not a
projector). For times greater than $t + \delta t$, there is no constraint, so the amplitude evolves
unitarily until some final moment $T$.

At the limit of small $\delta t$, the
successive operations above yield
\begin{eqnarray}
|\psi_0; \lambda, [t, t+ \delta t] \rangle =  - i \, \delta t \,
\,e^{-i\hat{H}(T - t)} \hat{P}_{\lambda} \hat{H} \hat{S}_t |\psi_0
\rangle. \label{amp1}
\end{eqnarray}

We emphasize here the important physical distinction on the role of time that is manifested in the derivation of the amplitude Eq.
(\ref{amp1}). The time of detection $t$ does {\em not} coincide with
the evolution parameter of Schr\"odinger' s equation. Instead, it is
a {\em physical observable} that corresponds to the moment that an {\em event} has taken place. The construction of
the amplitude $|\psi; \lambda, [t, t + \delta t] \rangle$ above
highlights this distinction: the detection time $t$ is {\em distinct} from
the  time $T$ at which the amplitude is evaluated.

\medskip

 The amplitude $|\psi_0; \lambda, [t, t + \delta t] \rangle$ is proportional to $\delta t$, hence it defines  a {\em density} with respect to time: $|\psi_0;  \lambda, t \rangle := \lim_{\delta t \rightarrow 0}
\frac{1}{\delta t} | \psi_0; \lambda, [t, t + \delta t] \rangle$. From Eq. (\ref{amp1})
\begin{eqnarray}
|\psi_0;  \lambda, t \rangle = - i   \,e^{-i\hat{H}(T - t)}
\hat{P}_{\lambda} \hat{H} \hat{S}_t |\psi_0 \rangle = - i e^{- i
\hat{H} T} \hat{C}(\lambda, t) |\psi_0 \rangle, \label{amp2}
\end{eqnarray}
where the class operator $\hat{C}(\lambda, t)$ is defined as
\begin{eqnarray}
\hat{C}(\lambda, t) = e^{i \hat{H}t} \hat{P}_{\lambda} \hat{H}
\hat{S}_t. \label{class}
\end{eqnarray}

 Since the amplitude $|\psi_0;  \lambda, t \rangle $ is a density
with respect to the time of transition $t$, its integration with respect to $t$ is well-defined. Hence, the total amplitude that the transition occurred at {\em some time} in the interval $[t_1,
t_2]$ is

\begin{eqnarray}
| \psi; \lambda, [t_1, t_2] \rangle = - i e^{- i \hat{H}T}
\int_{t_1}^{t_2} d t \hat{C}(\lambda, t) |\psi_0 \rangle.
\label{ampl5}
\end{eqnarray}

We note that if $[\hat{P}, \hat{H}] = 0$, i.e., if the Hamiltonian
evolution preserves the subspaces ${\cal H}_{\pm}$, then $|\psi_0;
\lambda, t \rangle = 0$. For a Hamiltonian    of the form $\hat{H} = \hat{H_0} + \hat{H_I}$, where $[\hat{H}_0, \hat{P}] = 0$, and $H_I$ a perturbing interaction, we obtain to leading order in the perturbation
\begin{eqnarray}
\hat{C}(\lambda, t) = e^{i \hat{H}_0t} \hat{P}_{\lambda} \hat{H}_I
e^{-i \hat{H}_0t}. \label{perturbed}
\end{eqnarray}
 Eq. (\ref{perturbed}) also holds for a time-dependent interaction Hamiltonian $\hat{H}_I$.

\subsection{Probabilities for detection time}

The amplitude  Eq. (\ref{amp2}) squared defines  the probability $p (\lambda, [t_1, t_2])\/$
that at some time in the interval $[t_1, t_2]$ a detection with
outcome $\lambda$ occurred
\begin{eqnarray}
p(\lambda, [t_1, t_2]) := \langle \psi; \lambda, [t_1, t_2] | \psi;
\lambda, [t_1, t_2] \rangle =   \int_{t_1}^{t_2} \,  dt
\, \int_{t_1}^{t_2} dt' \; Tr (e^{i\hat{H}( t - t')}
\hat{P}_{\lambda} \hat{H} \hat{S}^{\dagger}_t \hat{\rho}_0
\hat{S}_{t'} \hat{H} \hat{P}_{\lambda} ), \label{prob1}
\end{eqnarray}
where $\hat{\rho}_0 = |\psi_0\rangle \langle \psi_0|$.

However, the expression $p(\lambda, [t_1, t_2])$ does not correspond to a well-defined probability measure, because it
fails to satisfy the additivity condition. To see this, consider the probability corresponding to an
 interval $[t_1, t_3] = [t_1, t_2] \cup [t_2,
t_3]$. This equals
\begin{eqnarray}
p(\lambda, [t_1, t_3]) = p(\lambda, [t_1, t_2]) + p(\lambda, [t_2,
t_3])
+ 2 Re \left[ \int_{t_1}^{t_2} \,  dt \, \int_{t_2}^{t_3} dt'
Tr\left(\hat{C}(\lambda, t) \hat{\rho_0}\hat{C}^{\dagger}(\lambda,
t')\right)\right]. \label{add}
\end{eqnarray}
Hence, the additivity condition $p(\lambda, [t_1, t_3]) = p(\lambda, [t_1, t_2]) + p(\lambda, [t_2,
t_3])$, necessary for a consistent definition of a probability measure, fails, unless

\begin{eqnarray}
2 Re \left[
\int_{t_1}^{t_2} \,  dt \, \int_{t_2}^{t_3} dt'
Tr\left(\hat{C}(\lambda, t) \hat{\rho_0}\hat{C}^{\dagger}(\lambda,
t')\right)\right] = 0 \label{decond}
 \end{eqnarray}
 In the consistent/decoherent histories framework, Eq. (\ref{decond}) is referred to as the
 {\em consistency condition}. It is the minimal condition necessary for the association of a consistent probability measure in histories. It appears in the present framework, because we consider properties of the system at different moments of time, that is, histories.

   Eq. (\ref{decond}) does not hold for generic choices of $t_1, t_2$ and $t_3$. However, in a macroscopic system (or in a system with a macroscopic component) one expects that Eq. (\ref{decond}) holds with a good degree of approximation, given a sufficient degree of coarse-graining. This means that  there exists a
time-scale $\sigma$, such that the non-additive terms in Eq. (\ref{add}) are strongly suppressed if $ |t_2 - t_1| >> \sigma$ and $|t_3 - t_2| >> \sigma$. Then, Eq. (\ref{prob1}) does define a probability measure when restricted to intervals of size  larger than $\sigma$.

In principle, the coarse-graining time-scale $\sigma$ can be estimated from the knowledge of the Hamiltonian and the initial state of the system. Its value depends strongly on the detailed physics of the measurement scheme, and constitutes an inherent feature
 of a measurement process. Using a semi-classical language, $\sigma$ can be said to represent the time necessary for  a microscopic process to amplify and leave a definite macroscopic record. Thus, $\sigma$ places an absolute lower bound in the temporal resolution of each measurement scheme: it is impossible to distinguish events if their separation in time is of order $\sigma$ or less.

\paragraph{Temporal smearing.}
Assuming a finite coarse-graining time-scale $\sigma$, such that Eq. (\ref{decond}) is approximately valid for $ |t_2 - t_1| >> \sigma$ and $|t_3 - t_2| >> \sigma$, Eq. (\ref{prob1}) provides a consistent definition of a probability measure. However, it is more convenient to proceed by smearing the amplitudes
 Eq. (\ref{amp2}) at a time-scale of order $\sigma$ rather than employing probabilities for sharply defined time-intervals, as in Eq. (\ref{prob1}). Smearing allows us to express the probabilities in terms of densities
  with respect to a continuous variable.

To this end, we introduce a family of functions $f_{\sigma}(s)$,  localized around $s = 0$ with width $\sigma$, and normalized so that
$\lim_{\tau \rightarrow 0} f_{\tau}(s) = \delta(s)$. For example, one may employ the Gaussians
\begin{eqnarray}
f_{\sigma}(s) = \frac{1}{\sqrt{2 \pi \sigma^2}}
e^{-\frac{s^2}{2\sigma^2}}. \label{gauss}
\end{eqnarray}
The Gaussians Eq. (\ref{gauss}) satisfy the following equalities.
\begin{eqnarray}
f_{\sigma}(s) f_{\sigma}(s') = f_{\sigma}\left(\frac{s+s'}{2}\right) f_{\sigma}(s-s') \label{eq1}
\\
\sqrt{f_{\sigma}(t-s) f_{\sigma}(t-s')} = f_{\sigma}(t - \frac{s+s'}{2}) g_{\sigma}(s-s'), \label{eq2}
\end{eqnarray}
where
\begin{eqnarray}
g_{\sigma}(s) = \exp[-s^2/(8\sigma^2)]. \label{gsig}
 \end{eqnarray}
Eqs. (\ref{eq1}-\ref{eq2}) are satisfied approximately for other choices for the smearing functions with an error of order $\sigma^2$.

Using the functions $f_{\sigma}$, we  define the smeared amplitude $|\psi_0; \lambda, t\rangle_{\sigma}$ that is localized
 around the time $t$ with width $\sigma$, as
\begin{eqnarray}
|\psi_0; \lambda, t\rangle_{\sigma} := \int ds \sqrt{f_{\sigma}(s -t)}
|\psi_0; \lambda, s \rangle = \hat{C}_{\sigma}(\lambda, t) |\psi_0
\rangle, \label{smearing}
\end{eqnarray}
where
\begin{eqnarray}
\hat{C}_{\sigma}(\lambda, t) := \int ds \sqrt{f_{\sigma}(t - s)}
\hat{C}(\lambda, s).
\end{eqnarray}
The square amplitudes
\begin{eqnarray}
p(\lambda, t) = {}_{\sigma}\langle \psi_0; \lambda, t|\psi_0;
\lambda, t\rangle_{\sigma} = Tr \left[\hat{C}^{\dagger}_{\sigma}(\lambda, t)
 \hat{\rho}_0 \hat{C}_{\sigma}(\lambda, t)\right] \label{ampl6}
\end{eqnarray}
provide a well-defined probability measure: they are of the form
 $Tr[\hat{\rho}_0 \hat{\Pi}(\lambda, t)]$,
where
\begin{eqnarray}
\hat{\Pi}(\lambda, t) = \hat{C}_{\sigma}(\lambda, t)
\hat{C}^{\dagger}_{\sigma}(\lambda, t) \label{povm2}
\end{eqnarray}
is a density with respect to both variables $\lambda$ and $t$.

The positive
operator
\begin{eqnarray}
\hat{\Pi}_{\tau}(N) = 1 - \int_0^{\infty} dt \int d \lambda
\hat{\Pi}_{\tau}(\lambda, t), \label{nodet}
\end{eqnarray}
 corresponds to the alternative ${\cal N}$ that no detection took place
in the time interval $[0, \infty)$. $\hat{\Pi}_{\tau}(N)$ together with the positive operators Eq. (\ref{povm2})
 define a POVM on $ \left([0, \infty) \times
\Omega \right) \cup \{{\cal N} \}$, where $\Omega$ is the space of possible values of
$\lambda$. The POVM Eq. (\ref{povm2}) determines the probability density that a transition took place at time $t$, and that the outcome
$\lambda$ for the value of an observable has been recorded.

\subsection{Multi-time joint probabilities}

Next, we generalize the results above, by constructing the {\em joint probability} for transitions that occur at different moments of time.

 First, we consider the case of two detection events at times $t_1$ and $t_2$ ($t_1 < t_2$). Again, we denote the system's Hilbert space as ${\cal H}$.
 Let the transition at time $t_i$ ($i=1, 2$) correspond to a splitting of ${\cal H}$ into two orthogonal subspaces associated to the projectors $\hat{P}_{i+}$ and $\hat{P}_{i-} = 1 - \hat{P}_{i+}$. At each time $t_i$, the values $\lambda_i$ are being recorded; these values correspond to an exclusive set of projectors $\hat{P}_{\lambda_i}$, such that $\sum_{\lambda_i} \hat{P}_{\lambda_i} = \hat{P}_{i+}$.

The probability amplitudes $|\psi_0;\lambda_1,[t_1, t_1 + \delta t_1]; \lambda_2, [t_2, t_2+\delta t_2] \rangle$ that correspond to the recording of values $\lambda_i$  at times $t_i$ are obtained through a similar procedure to the one leading to Eq. (\ref{amp1}):
\begin{eqnarray}
|\psi_0;\lambda_1,[t_1, t_1 + \delta t_1]; \lambda_2, [t_2, t_2+\delta t_2] \rangle = - \delta t_1 \delta t_2 e^{-i \hat{H}T} \hat{C}(\lambda_1, t_1; \lambda_2, t_2) | \psi_0\rangle, \label{psit12}
\end{eqnarray}
where
\begin{eqnarray}
\hat{C}(\lambda_1, t_1; \lambda_2, t_2) := e^{i \hat{H}t_2} \hat{P}_{\lambda_2} \hat{S}^{(2)}_{t_2-t_1} \hat{P}_{\lambda_1} \hat{H} S_{t_1}^{(12)}. \label{c12}
\end{eqnarray}

In Eq. (\ref{c12}), $S_t^{(12)}$ is the restricted propagator corresponding to no transitions (either with respect to $\hat{P}_{1+}$ or with respect to $\hat{P}_{2+}$), that is,
\begin{eqnarray}
\hat{S}_t^{(12)} = \lim_{n \rightarrow \infty} \left[ \hat{P}_{12-} e^{-i\hat{H}t/n} \hat{P}_{12-} \right]^n,
\end{eqnarray}
where $\hat{p}_{12-}$ is the projector corresponding to the logical disjunction of $\hat{P}_{1-}$ and $\hat{P}_{2-}$. For $[\hat{P}_{1-}, \hat{P}_{2-}] = 0$, which is the case of interest in this paper (it corresponds to independent apparatuses), $\hat{P}_{12-} = \hat{P}_{1-} \hat{P}_{2-}$. A similar expression defines $\hat{S}^{(2)}_t$, the restricted propagator corresponding to a transition with respect to $\hat{P}_{1+}$, but not with respect to $\hat{P}_{2+}$.

The explicit form of the restricted propagators is not important when the  Hamiltonian $\hat{H}$ is of the form $\hat{H} = \hat{H}_0 + \hat{H}_I$, where $[\hat{H}_0, \hat{P}_{i+}] = 0$ and $\hat{H}_I$ is a small perturbation. In this case, the class operator $\hat{C}(\lambda_1, t_1; \lambda_2, t_2)$ of Eq. (\ref{c12}) becomes
\begin{eqnarray}
\hat{C}(\lambda_1, t_1; \lambda_2, t_2) = e^{i \hat{H}_0t_2} \hat{P}_{\lambda_2} \hat{H}_I e^{- i \hat{H}_0(t_2 - t_1)} \hat{P}_{\lambda_1} e^{-i \hat{H}_0 t_1}, \label{c12b}
\end{eqnarray}
to leading order in perturbation theory.

To obtain the associated probability density, we smear the amplitude (\ref{psit12}) at each moment of time $t_1$ and $t_2$, using smearing functions $f_{\sigma_1}$ and $f_{\sigma_2}$, respectively. In general, the coarse-graining time-scales $\sigma_i$ may be different for  each transition. Following the same procedure as in Sec. 2.2, we obtain

\begin{eqnarray}
p(\lambda_1, t_2; \lambda_2, t_2) = \int ds_1 ds_1' ds_2 ds_2' \sqrt{f_{\sigma_1}(t_1-s_1)f_{\sigma_1}(t_1-s_1') f_{\sigma_2}(t_2-s_2) f_{\sigma_2}(t_2 - s_2')}\nonumber \\
\times \langle \psi_0| \hat{C}^{\dagger}(\lambda_1, s_1;\lambda_2, s_2) \hat{C}(\lambda_1, s_1';\lambda_2, s_2')| \psi_0\rangle. \label{p12}
\end{eqnarray}

The procedure above can be applied systematically in order to obtain the $n$-time joint probabilities $p(\lambda_1, t_1; \lambda_2, t_2; \ldots ; \lambda_n, t_n)$ that correspond to measurements of observables $\lambda_1, \lambda_2, \ldots, \lambda_n$ at times $t_1, t_2, \ldots, t_n$ respectively ($t_1 < t_2 < \ldots < t_n$). Here, we give the final expression for a Hamiltonian $\hat{H}_0 + \hat{H}_I$, where $[\hat{H}_0, \hat{P}_{i+}] = 0$ and $\hat{H}_I$ is a small perturbation,  to leading order in perturbation theory
\begin{eqnarray}
p(\lambda_1, t_1; \lambda_2, t_2; \ldots ; \lambda_n, t_n) = \int \prod_{i = 1}^n ds_i ds_i' \sqrt{f_{\sigma_i}(t_i - s_i) f_{\sigma_i}(t_i - s_i')} \nonumber \\
\times \langle \psi_0|\prod_{j = 1}^n \left( e^{i \hat{H}_0 s_j'}\hat{P}_{\lambda_j} \hat{H}_I e^{- i \hat{H}_0 s_j'} \right) \prod_{k = 1}^n \left( e^{i \hat{H}_0 s_k} \hat{P}_{\lambda_k} \hat{H}_I e^{-i\hat{H}_0 s_k}| \psi_0 \rangle\right). \label{22}
\end{eqnarray}

\subsection{A general model for particle detectors}

So far, we have constructed expressions for the single-time and  multi-time probabilities for a generic quantum system. In what follows, we  study the case of an apparatus that responds to the excitations of a microscopic system and records their energy. We obtain expressions for the intensity of the absorbed energy as a function of time and  for the intensity fluctuations.

\subsubsection{Intensity at a single moment of time}

 We consider a Hilbert space  ${\cal H}_{det}$, describing the degrees of freedom of a macroscopic detector, and a Hilbert space  ${\cal H}_q$, describing a microscopic quantum system interacting with the detector. The Hilbert space of the combined system is then ${\cal H} = {\cal H}_{det} \otimes {\cal H}_q$. We denote the Hamiltonian of the detector degrees of freedom as $\hat{H}_d$ and the Hamiltonian of the measured system as $\hat{H}_q$. A general form of the interaction Hamiltonian is $\sum_a \hat{A}^a \otimes \hat{B}^a_t$, where $\hat{A}^a$ are operators on ${\cal H}_{det}$ and $\hat{B}^a_t$ (possibly time-dependent) operators on ${\cal H}_q$.

  Our aim is to model  a detector that records the energy corresponding to the excitations of the quantum system. The relevant transitions correspond to changes in the detector's energy, as it absorbs an excitation (particle) of the microscopic system. We assume that the initially prepared state of the detector has minimum energy $E_0$, conveniently taken equal to zero, and that the excited states have energies $E > E_0$. We denote the projectors onto the constant energy subspaces as $\hat{P}_E$. Thus, the projectors $\hat{P}_{\pm}$ on ${\cal H}$ corresponding to the transitions are $\hat{P}_+ = \left(\sum_{E>E_0} \hat{P}_E \right) \otimes \hat{1}$ and $\hat{P}_- = \hat{P}_0 \otimes \hat{1}$, where  $\hat{P}_E = \hat{P}_{E_0}$.

The class operators Eq. (\ref{perturbed}) for this system are
\begin{eqnarray}
\hat{C}(E,t) = \sum_a \left(\hat{P}_E \hat{A}^a(t)\right) \otimes \hat{B}^a(t),
\end{eqnarray}
where $\hat{A}^a(t) = e^{i \hat{H}_d t} \hat{A}^a e^{-i\hat{H}_dt}$ and $\hat{B}^a(t) = e^{i \hat{H}_qt} \hat{B}^a_t e^{- i \hat{H}_qt}$ are the Heisenberg-picture evolution of $\hat{A}^a$ and $\hat{B}^a_t$ respectively.

Physically reasonable detectors are characterized by an energy gap, i.e, there exists a minimal value $E_{min} > E_0$ in the energies of the excited states. If the detector's temperature is much smaller than $E_{min}$, then a thermal state for the detector' degrees of freedom is well approximated by the density matrix $\hat{\rho}^{(0)}_{det} = \hat{P}_0/Tr\hat{P}_0$.

We consider an initial state of the total system $\hat{\rho}^{(0)}\otimes \hat{\rho}_0$, where $\hat{\rho}_0$ is the initial state of the microscopic system. Eq. (\ref{ampl6}) then gives
\begin{eqnarray}
p(E,t) = \sum_{a,b} \alpha^{ab}(E) \int ds \int ds' \sqrt{f_{\sigma}(t-s) f_{\sigma}(t-s')} e^{-iE(s'-s)} Tr_{{\cal H}_q} \left[\hat{\rho}_0 \hat{B}^a(s') \hat{B}^b(s)\right], \label{p1b}
\end{eqnarray}
where
\begin{eqnarray}
\alpha^{ab} = \frac{Tr_{{\cal H}_{det}} \left( \hat{P}_0 \hat{A}^a \hat{P}_E \hat{A}^b\right)}{Tr_{{\cal H}_{det}} \hat{P}_0}.
\end{eqnarray}

The intensity $I$ recorded by the detector at time $t$ (energy absorbed per unit time) is then obtained from Eq. (\ref{p1b})
\begin{eqnarray}
I(t) = \int dE E p(E,t). \label{I}
\end{eqnarray}

\subsubsection{Intensity fluctuations}
Next, we consider the intensity fluctuations measured by one or more detectors. To this end,
we assume that the detector, being macroscopic, consists of a large number of independent subsystems, each of which being able to record a microscopic excitation. The intensity refers to the totality of detection events, irrespective of the subsystem where they have been recorded. Hence, in order to construct the $n$-time intensity correlation function, we must consider the response of $n$ independent detectors, identical to the ones considered in Sec. 2.3.

We first work for $n=2$; the generalization to higher $n$ is straightforward. We consider two copies of a detector's Hilbert space ${\cal H}_{det}$, so that the Hilbert space of the total system is ${\cal H}_{det1}\otimes {\cal H}_{det2} \otimes {\cal H}_q$. The projector $\hat{P}_{+1}$ corresponding to the transition recorded by  detector $1$ is $\sum_{E>E_0} \hat{P}_E\otimes 1\otimes 1$ and the projector $\hat{P}_{+2}$ corresponding to the transition recorded by detector $2$ is $1\otimes \sum_{E>E_0} \hat{P}_E\otimes 1$.

We assume that there is no direct interaction between the two detectors. The general form of the interaction Hamiltonian is then
\begin{eqnarray}
\hat{H}_I = \sum_a \left(\hat{A}^a\otimes \hat{1} \otimes \hat{B}^a_t + \hat{1} \otimes \hat{A}^a \otimes \hat{B}^a_t \right).
\end{eqnarray}

Substituting into Eq. (\ref{c12b}) we obtain the following expression for the class operator $\hat{C}(E_1, t_1, E_2, t_2)$

\begin{eqnarray}
\hat{C}(E_1, t_1, E_2, t_2) = \sum_{a,b} \left( \hat{P}_{E_2} \hat{A}^a(t_1) \right) \otimes \left( \hat{P}_{E_2} \hat{A}^b(t_2)\right) \otimes \left( \hat{B}^b(t_2) \hat{B}^a(t_1)\right). \label{c12c}
\end{eqnarray}

In the derivation of Eq. (\ref{c12c}), we assumed that the first detection (at time $t_1$) was made by the detector $1$ and the second detection (at time $t_2$) by the detector 2. If the time-ordering of the two detections is not observable, we must also take into account a contribution to the amplitude with the reverse time-ordering for the detections. Then, the relevant class operator is
\begin{eqnarray}
\hat{C}(E_1, t_1, E_2, t_2) = \sum_{a,b} \left( \hat{P}_{E_2} \hat{A}^a(t_1) \right) \otimes \left( \hat{P}_{E_2} \hat{A}^b(t_2)\right) \otimes {\cal T} \left( \hat{B}^b(t_2) \hat{B}^a(t_1)\right),
\end{eqnarray}
where ${\cal T}$ stands for time-ordering.

Thus, from Eq. (\ref{p12}) we find the probability density $p(E_1,t_1;E_2, t_2)$ for a detection with energy $E_1$ at time $t_1$ and a detection with energy $E_2$ at time $t_2$
\begin{eqnarray}
p(E_1, t_1, E_2, t_2) &=& \sum_{a,b,a',b'} \alpha^{a'a}(E_1) \alpha^{b'b}(E_2) \int ds_1 ds_1' ds_2 ds_2'
e^{-iE_1(s_1'-s_1) - iE_2(s_2'-s_2)}
\nonumber \\
&\times&
  \sqrt{f_{\sigma_1}(t_1-s_1)f_{\sigma_1}(t_1-s_1') f_{\sigma_2}(t_2-s_2) f_{\sigma_2}(t_2 - s_2')} \nonumber \\
&\times&  Tr_{{\cal H}_q} \left[ \hat{\rho}_0 {\cal A} \left( \hat{B}^{a'}(s_1') \hat{B}^{b'}(s_2')\right) {\cal T} \left(\hat{B}^b(s_2) \hat{B}^a(s_1)\right) \right], \label{pd12}
\end{eqnarray}
where ${\cal A}$ stands for the anti-time-ordered product.

The two-time intensity correlation function is defined as
\begin{eqnarray}
\langle I(t_1)I(t_2)\rangle := \int dE_1 dE_2 E_1 E_1 p(E_1, t_1;E_2,t_2). \label{II}
\end{eqnarray}

We similarly derive the $n$-time detection probability density
\begin{eqnarray}
p(E_1, t_1; E_2, t_2; \ldots; E_n, t_n)&{}& = \sum_{a_1, \ldots, a_n} \sum_{a_1', \ldots, a_n'} \prod_{i=1}^n\left( \alpha^{a_i a_i'}(E_i) \right.
\nonumber \\
&\times& \left. \int  ds_i ds_i' e^{-iE_i(s_i'-s_i)}
 \sqrt{f_{\sigma_i}(t_i-s_i) f_{\sigma}(t_i-s_i')} \right.
\nonumber \\
&\times& \left.
Tr_{{\cal H}_q } \left[ \hat{\rho}_0 {\cal A} \left(\hat{B}^{a_1'}(s_1') \ldots \hat{B}^{a_n'}(s_n')\right){\cal T} \left(\hat{B}^{a_n}(s_n) \ldots \hat{B}^{a_1}(s_1)\right)\right]\right), \hspace{0.5cm} \label{ptn}
\end{eqnarray}
and the $n$-time intensity correlation function
\begin{eqnarray}
\langle I(t_1) \ldots I(t_n) \rangle = \int dE_1 \ldots dE_n E_1 \ldots E_n p(E_1, t_1; E_2, t_2; \ldots; E_n, t_n).
\end{eqnarray}

\section{The Unruh-DeWitt detector}

In Sec. 2, we derived the equations for the joint probabilities and intensity correlations corresponding to general macroscopic detectors interacting with a quantum system. In this section, we specialize to detectors of the  Unruh-DeWitt type that move along general trajectories in Minkowski spacetime and interact with a massless scalar field.

\subsection{Detection probability for Unruh-DeWitt detectors}

\subsubsection{Derivation of a general expression}

An Unruh-DeWitt detector is an ideal particle detector that interacts  with a quantum field via a dipole coupling. One usually considers such detectors moving along non-inertial paths in Minkowski spacetime. In this context, a crucial assumption is that the physics at the detector's rest frame is independent of the path followed by the detector \cite{Pad0}. Hence,  the evolution operator for an Unruh-DeWitt detector is   $e^{-i\hat{H}_d\tau}$, where $\tau$ is the proper time along the detector's path and $\hat{H}_d$ is the Hamiltonian for a stationary detector.  Thus the path-dependence enters into the propagator only through the proper time. In particular, supposing that the detector follows a timelike path $x^{\mu}(\tau)$ in Minkowski spacetime, the equation $x^0(\tau) = t$ can be solved in order to determine the proper time $\tau(t)$ as a function of the inertial time coordinate $t$ of Minkowski spacetime. Then the evolution operator for the detector with respect to $t$ is
$e^{-i\hat{H}_d\tau(t)}$.

We assume that the  detector is coupled to a massless scalar field $\hat{\phi}$.
The scalar field Hamiltonian is $\hat{H}_{\phi} = \int d^3 x \left( \frac{1}{2} \hat{\pi}^2 + \frac{1}{2} (\nabla \phi)^2\right)$. The evolution operator for the field with respect to a Minkowskian inertial time coordinate $t$  is $e^{- i \hat{H}_{\phi}t}$, where $t$ is a Minkowski inertial time parameter. The Unruh-DeWitt detector interacts with the field through a dipole coupling. This means that the (time-dependent)  interaction Hamiltonian is of the form $\hat{H}_I = \hat{m} \otimes \hat{\phi}[x(\tau)]$, where $\hat{\phi}(x)$ are the Heiseberg-picture fields, and   $\hat{m}$ is an operator on the detector's Hilbert space analogous to the magnetic moment in electrodynamics.

 The detection probability for an Unruh-DeWitt detector follows from Eq. (\ref{p1b}), with the substitutions $\hat{A}^a \rightarrow \hat{m}$, $\hat{B}^a(t) \rightarrow \hat{\phi}[x(\tau)]$.  Using the detector's proper time $\tau$ as a time parameter, we obtain

\begin{eqnarray}
p(E, \tau) = \alpha(E) \int ds \int ds' \sqrt{f_{\sigma}(t-s) f_{\sigma}(t-s')} e^{-iE(s'-s)} \Delta^+[x(s'),x(s)], \label{pet}
\end{eqnarray}
where $\alpha(E) = Tr_{{\cal H}_d}(\hat{P}_E\hat{m} \hat{P}_0\hat{m})/Tr_{{\cal H}_d} \hat{P}_0$, and
\begin{eqnarray}
\Delta^+(x,x') = Tr\left[ \hat{\phi}(x) \hat{\phi}(x') \hat{\rho}_0 \right]
\end{eqnarray}
is the positive-frequency Wightman function. When the field is in the vacuum state,
\begin{eqnarray}
\Delta^+(x,x') = \frac{-1}{4\pi^2 [(x^0-x^{0'}-i \epsilon)^2 - ({\bf x} - {\bf x'})^2]},
\end{eqnarray}
 where $\epsilon >0$ is the usual regularization parameter.

Eq. (\ref{pet}) simplifies using Eq. (\ref{eq1}) and then setting $f_{\sigma}(\tau - \frac{s+s'}{2}) \simeq \delta(\tau - \frac{s+s'}{2})$. This is a good approximation when the system is monitored at timescales much larger than $\sigma$. Then,
\begin{eqnarray}
p(E, \tau) = \alpha(E) \int dy g_{\sigma}(y) e^{-iEy} \Delta^+[x(\tau+\frac{y}{2}), x(\tau - \frac{y}{2})]. \label{petb}
\end{eqnarray}
 Eq. (\ref{petb}) determines the detector's response for motion along a general spacetime path. It
 is similar to the standard expression obtained through first-order perturbation theory with an adiabatic switching on and off of the coupling \cite{Schlicht, Lang, LouSa}.
The difference lies on the factor $g_{\sigma}(y)$. This term guarantees that the detector's response to changes in its state of motion is causal and approximately local in time at macroscopic scales of observation. To see this, we note that $g_{\sigma}(y)$ truncates contributions to the detection probability from all instants $s$ and $s'$ such that $|s-s'|$ is substantially larger than $\sigma$. Thus, at each moment of time $\tau$, the detector's response is determined solely from properties of the path at times around $\tau$ with a width of order $\sigma$. In particular, properties of the path at the asymptotic past (or future) do not affect the detector's response at time $\tau$. The response is determined solely by the properties of the path at time $\tau$, within the accuracy allowed by the detector's temporal resolution.

 At the limit $\sigma \rightarrow \infty$, $g_{\sigma}(y) \rightarrow 1$, and the probability $p(E, t)$ is non-local in time and non-causal. According to the discussion in Sec. 2.2, this limit corresponds to a physical system for which the detection event is not be localized in time.
 A finite value of $\sigma$ is indicative of a detector with many degrees of freedom, and thus the records of particle detection are coarse-grained observables. Coarse-graining typically implies noise that may deform the `signal' due to the detector's state of motion. We will see that this is indeed the case for uniform accelerated detectors, where a finite value of $\sigma$ implies a deformation of the Planckian spectrum.

 \subsubsection{Uniformly accelerated detectors}

 Next, we apply Eq. (\ref{pet}), in order to obtain the response of a uniformly accelerated detector. We consider a spacetime trajectory of the form
  $x^0(\tau) = \sinh(a\tau)/a, x^1(\tau) = (\cos(a\tau) - 1)/a, x^2(\tau) = x^3(\tau) = 0$. Then,
\begin{eqnarray}
p(E, \tau) = - \frac{\alpha(E) a^2}{16 \pi^2} \int \frac{dy g_{\sigma}(y) e^{-iEy}}{\sinh^2[(ay - i \tilde{\epsilon})/2]}, \label{petunr}
\end{eqnarray}
where a time-dependent factor has been absorbed into $\tilde{\epsilon}$.

The integrand in Eq. (\ref{petunr}) is peaked around $y = 0$. The hyperbolic sinus in the denominator has a width of order $2/a$ around $y = 0$, and the function $g_{\sigma}$ has a peak of width $\sigma$ around $y= 0$. If $\sigma a >> 1$ and  $E/a$ is not much smaller than unity, we  obtain $p(E, \tau)$ to leading order in $1/(\sigma a)$ by setting $g_{\sigma} = 1$. We find
\begin{eqnarray}
p(E, \tau) \simeq - \frac{\alpha(E) a^2}{16 \pi^2} \int  \frac{dy e^{-iEy}}{\sinh^2[a(y - i \tilde{\epsilon})/2]} = \frac{\alpha(E)}{2 \pi} \frac{E}{e^{\frac{2\pi E}{a}} - 1},
\end{eqnarray}
corresponding to a Planckian spectrum for the distribution of energy, with temperature $T = \frac{a}{2\pi}$.

 To compute corrections due to the  finite value of  $1/(\sigma a)$, we use Eq. (\ref{gsig}) for $g_{\sigma}$ and express Eq. (\ref{petunr}) as
\begin{eqnarray}
p(E, \tau) = - \frac{\alpha(E) a}{16 \pi^2} \int dz \frac{e^{-\eta z^2/8 - i Ez/a}}{\sinh^2[z-i \tilde{\epsilon})/2]}, \label{pexp}
\end{eqnarray}
where $\eta = (\sigma a)^{-2}$ and $z = ay$.

 We evaluate the integral Eq. (\ref{pexp}) as a power series in $\eta$, to obtain
\begin{eqnarray}
p(E, \tau) = \frac{\alpha(E)}{2 \pi} \frac{E}{e^{\frac{2\pi E}{a}} - 1} \left[ 1 + \frac{1}{\sigma^2 a^2}\left(\frac{\pi^2}{4} \frac{e^{\frac{2\pi E}{a}} +1}{(e^{\frac{2\pi E}{a}} - 1)^2} - \frac{\pi a}{4E(1 -e^{\frac{2\pi E}{a}})} \right) + \ldots \right].
\end{eqnarray}
Hence, provided that $a/E$ is not much larger than unity, the effect of a finite value for $1/(\sigma a)$ is a small deviation from the Planckian spectrum. However, when the condition  $\sigma a >> 1$ is not satisfied, the response function takes an entirely different form and has no similarity to thermal radiation. We note that the condition $a \sigma >> 1$ for the identification of the Unruh  temperature is mirrored in the study of the Unruh effect for observers with finite lifetimes $\tau_0$ \cite{rov, mar}. The physical context in the above references is rather different, but they confirm the point that the derivation of the Unruh temperature requires that the time-scale $1/a$ associated to the acceleration is much smaller than any intrinsic time-scales of the detecting system.

 We conclude that, in a macroscopic detector, Unruh's relation between acceleration and temperature requires a separation of timescales. If the resolution timescale is of the order of   $1/a$, the quantum coherence of the microscopic processes related to particle detection is lost. However, even if the resolution timescale $\sigma$ is much larger than $1/a$, it may still be much smaller than the macroscopic timescales characterizing the detector's trajectory. For example,  consider a path characterized by a stage of uniform acceleration of proper-time duration equal to $\Delta \tau >> \sigma$. During this period, the detector's response will be Planckian, modulo transient effects of order $\sigma/\Delta \tau$ and spectrum deformations of order $(\sigma a)^{-2}$. Thus the local-in-time thermal response of an accelerated detector arises in this calculation as a consequence of the  separation of three time-scales: the microscopic scale $1/a$, the resolution scale $\sigma$ and the macroscopic scale $\Delta \tau$ characterizing variations in the detector's acceleration.

We next compare the response of the accelerated detector in the Minkowski vacuum, with a response of static detector ($x^0 = t, x^i - 0$) in a thermal bath at temperature $T = \beta^{-1}$. The corresponding positive-frequency Wightman function for a massless quantum field  is
\begin{eqnarray}
\Delta^+_{\beta}(t, {\bf x}; t',  {\bf x'}) &=& -\frac{1}{8\pi\beta r} \frac{\sinh(2 \pi r/\beta)}{\sinh[\pi(t-t'-i \epsilon- |{\bf x}-{\bf x'}|)/\beta]\sinh[\pi(t-t'-i \epsilon+ |{\bf x}-{\bf x'}|)/\beta]} \nonumber  \\
&-& \frac{i}{4 \pi} \sum_{n = - \infty}^{\infty \prime} \delta (t -t' + i n \beta), \label{deltab}
\end{eqnarray}
where the prime denotes that the $n = 0 $ term is excluded from the summation. For ${\bf x} = {\bf x'} = 0$, Eq. (\ref{deltab}) becomes
\begin{eqnarray}
\Delta^+_{\beta}(t, t') = -\frac{-1}{4 \beta^2 \sinh^2[\pi(t - t' - i \epsilon)/\beta]}  - \frac{i}{4 \pi} \sum_{n = - \infty}^{\infty \prime} \delta (t -t' + i n \beta). \label{deltab2}
\end{eqnarray}

Substituting Eq. (\ref{deltab2}) into Eq. (\ref{petb}), we find that at the limit $\sigma a >> 1$ ($g_{\sigma} \simeq 1$, the delta functions do not contribute to the total integral because they vanish along the real axis. Hence, the response function $p(E, \tau)$ coincides, as expected,
with that of an accelerating detector for $T = \frac{a}{2\pi}$. This equivalence may be valid for state other than thermal. Any Wightman function with poles in the negative imaginary plane at $t - t' = - in \kappa$, where $n = 1, 2, \ldots$ and $\kappa$ a positive constant gives rise to a Planckian spectrum (for $\kappa \sigma >> 1$).

\subsubsection{Detector's response for other paths}

The response function Eq. (\ref{petb}) also applies to other choices of spacetime paths. We examine some cases:

\medskip

 1. For paths such that $\Delta^+[x(s), x(s')]$ is a function only of $s-s'$, the detection rate is constant. This result also follows from first-order perturbation theory in a microscopic detector \cite{Pad0}.

 \medskip

2. For paths with time-dependent linear acceleration $a(\tau)$, such that the acceleration varies at a scale much larger than $\sigma$ ($\dot{a}/a << 1/\sigma$), the detection rate corresponds to  a time-dependent Planckian spaectrum (as long as $a(\tau) \sigma >> 1$), i.e.,
        \begin{eqnarray}
        p(E, \tau) \simeq \frac{\alpha(E)}{2 \pi} \frac{E}{e^{\frac{2 \pi E}{a(\tau)}}-1}.
        \end{eqnarray}
Thus the detector comes to equilibrium at the instantaneous values of the Unruh temperature $T(\tau) = \frac{a(\tau)}{2 \pi}$.

\medskip

3. For paths of the form $(x^0(\tau), x^1(\tau), 0, 0)$ that correspond to motion along a single axis, we define $u(\tau) = x^0(\tau) + x^1(\tau)$ and $v(\tau) = x^0(\tau) - x^1(\tau)$. Clearly $\dot{u} \dot{v} = -1$. Eq. (\ref{petb}) becomes
    \begin{eqnarray}
    p(E, \tau) = - \frac{\alpha(E)}{4 \pi^2} \int dy \frac{g_{\sigma}(y) e^{- i Ey}}{[u(\tau+y/2) - u(\tau - y/2)-i \epsilon][v(\tau+y/2)-v(\tau - y/2) - i \epsilon]}. \label{uv}
    \end{eqnarray}

    The function $g_{\sigma}(y)$ effectively restricts integration to values of $y$ of order of $\sigma$ or smaller. Thus, we can Taylor-expand the terms in the denominator of Eq. (\ref{uv}) as $u(\tau +y/2) - u(\tau - y/2) = y \dot{u}(\tau) + \sum_{k =1}^{\infty} \frac{u^{(2k+1)}(\tau)}{2^k (2k+1)!} y^{2k}]$, and truncate the series to some order $k_m$ if $u^{(2k_m+3)}(\tau)\sigma^{2k}$ becomes significantly smaller than the previous terms in the expansion. Similarly, we can truncate the series for $u(\tau + y/2) - u(\tau - y/2)$ at some order $k_m'$, so that the denominator in Eq. (\ref{uv}) is approximated by a polynomial of order $k_m + k_m'$. We then obtain the dominant contribution to $p(E, \tau)$ by letting $g_{\sigma} \rightarrow 1$, whence the integral corresponds to the sum of the residues for the poles lying in the negative imaginary plane. For example, if we truncate the series at $k_m = k_m' = 1$, the integrand in Eq. (\ref{uv}) has a double pole at $y = 0$ and also poles at $y^2 = -24 \dot{u}/\dddot{u}$ and $y^2 = - 24 \dot{v}/\dddot{v}$. If $0 < \dot{u}/\dddot{u} \neq \dot{v}/\dddot{v} >0$, there are two single poles in the negative imaginary plane, whence

\begin{eqnarray}
p(E, \tau) = \frac{\alpha(E)}{4 \pi} \frac{\sqrt{\frac{24 \dddot{u}}{\dot{u}}} e^{- \sqrt{\frac{24 \dot{u}}{\dddot{u}}}E } - \sqrt{\frac{24 \dddot{v}}{\dot{v}}} e^{- \sqrt{\frac{24 \dot{v}}{\dddot{v}}}E }}{\dot{u} \dddot{v}-\dot{v} \dddot{u}}.
\end{eqnarray}

In contrast, if $\dot{u}/\dddot{u}, \dot{v}/\dddot{v}<0$, all poles are real-valued and thus $p(E, \tau)$ vanishes.

\subsection{Intensity fluctuations for the Unruh-DeWitt detector}

 \subsubsection{The general case}
 Next, we construct  the joint detection probability $p(E_1, \tau_1; E_2, \tau_2)$ for a pair of identical detectors that move along different spacetime trajectories, $x_1^{\mu}(\tau_1)$ and $x_2^{\mu}(\tau_2)$.  Eq. (\ref{pd12}) applies to this case. We substitute $\hat{A}^a \rightarrow \hat{m}$, $\hat{B}^a(t) \rightarrow \hat{\phi}[x_1(\tau_1)]$, $\hat{B}^b(t) \rightarrow \hat{\phi}[x_2(\tau_2)]$, and we use the proper-time parameters $\tau_1$ and $\tau_2$, corresponding to Minkowskian times $t_1$ and $t_2$ via $x^0_1(\tau_1) = t_1$ and $x^0_2(\tau_2) = t_2$. We obtain
 \begin{eqnarray}
 p(E_1, \tau_1; E_2, \tau_2) &=& \alpha(E_1) \alpha(E_2) \int ds_1 ds_1' ds_2 ds_2' e^{-iE_1(s_1' - s_1) - i E_2 (s_2' - s_2)}
 \nonumber \\
 &\times &
  \sqrt{f_{\sigma}(t_1-s_1)f_{\sigma}(t_1-s_1') f_{\sigma}(t_2-s_2) f_{\sigma}(t_2 - s_2')}
 \nonumber \\
 &\times &Tr_{{\cal H}_{\phi}} \left[ {\cal A} \left( \hat{\phi}[x_1(s_1')] \hat{\phi}[x_2(s_2')]\right) {\cal T} \left( \hat{\phi}[x_2(s_2)] \hat{\phi}[x_1(s_1)]\right) \hat{\rho}_0 \right], \label{petet}
 \end{eqnarray}
 where time- and anti-time-ordering refers to the Minkowski time parameters that are associated to the proper times $s_1, s_2$ and $s_1', s_2'$ respectively. When $\hat{\rho}_0$ is the Minkowski vacuum state, Eq. (\ref{petet}) becomes
 \begin{eqnarray}
  p(E_1, \tau_1; E_2, \tau_2) = \alpha(E_1) \alpha(E_2) \int ds_1 ds_1' ds_2 ds_2' e^{-iE_1(s_1' - s_1) - i E_2 (s_2' - s_2)}
  \nonumber \\
  \times
  \sqrt{f_{\sigma}(t_1-s_1)f_{\sigma}(t_1-s_1') f_{\sigma}(t_2-s_2) f_{\sigma}(t_2 - s_2')}
 \left( \Delta_F^*[x_2(s_2'), x_1(s_1')] \Delta_F[x_2(s_2), x_1(s_1)] \right.
 \nonumber \\
 \left. + \Delta^+[x_2(s_2'),x_2(s_2)] \Delta^+[x_1(s_1'),x_1(s_1)] +  \Delta^+[x_2(s_2'),x_1(s_1)] \Delta^+[x_1(s_1'),x_2(s_2)] \right), \label{petet2}
 \end{eqnarray}
 where $\Delta_F(x,x') = - 1/[4 \pi^2(x-x')^2] - \frac{i}{4 \pi} \delta[(x-x')^2]$ is the Feynman propagator for the scalar field. The second product of Green's function in Eq. (\ref{petet}) gives rise to $p(E_1, \tau_1) p(E_2, \tau_2)$. Thus,
 \begin{eqnarray}
 p(E_1, \tau_1; E_2, \tau_2) = p(E_1, \tau_1) p(E_2, \tau_2) + G(E_1, \tau_2; E_2, \tau_2), \label{pg}
 \end{eqnarray}
 where the function
 \begin{eqnarray}
  G(E_1, \tau_1; E_2, \tau_2) = \alpha(E_1) \alpha(E_2) \int ds_1 ds_1' ds_2 ds_2' e^{-iE_1(s_1' - s_1) - i E_2 (s_2' - s_2)} \hspace{2cm} \nonumber \\
  \times \sqrt{f_{\sigma}(t_1-s_1)f_{\sigma}(t_1-s_1') f_{\sigma}(t_2-s_2) f_{\sigma}(t_2 - s_2')}
 \left( \Delta_F^*[x_2(s_2'), x_1(s_1')] \Delta_F[x_2(s_2), x_1(s_1)]
  \right.
  \nonumber \\
\left.  + \Delta^+[x_2(s_2'),x_1(s_1)] \Delta^+[x_1(s_1'),x_2(s_2)] \right), \label{G}
 \end{eqnarray}
 reflects the correlation between the detection events in the two detectors.

Eqs. (\ref{pg}-\ref{G}) apply to general paths in Minkowski spacetime. They simplify significantly, when we restrict to paths $x_1(\tau)$ and $x_2(\tau)$,
 such that the Green's functions $\Delta_F[x_1(s),x_2(s')]$ and $\Delta[x_1(s),x_2(s')]$ are functions of $s-s'$ only. Denoting the values of the Green's functions values as $\Delta_F(s-s')$ and $\Delta^+(s-s')$ respectively, Eq. (\ref{G}) becomes
 \begin{eqnarray}
  G(E_1, \tau_1; E_2, \tau_2) = \alpha(E_1) \alpha(E_2) \int dS_1 dS_2 f_{\sigma}(\tau_1-S_1) f_{\sigma}(\tau_2 - S_2)
  \nonumber \\
 \times  \int dx dy e^{-i(E_1+E_2)x - i(E_1-E_2)y/2}
  g_{\sigma}(\sqrt{2}x) g_{\sigma}(y/\sqrt{2}) \hspace{3cm}
   \nonumber \\
   \times \left[\Delta_F^*(S_2-S_1+y)\Delta_F(S_2-S_1-y) + \Delta^+(S_2-S_1+x) \Delta^+(S_1-S_2+x)\right], \label{G2}
 \end{eqnarray}
 where $S_1= (s_1+s_1')/2$, $S_2 = (s_2+s_2')/2$, $x = \frac{1}{2}(s_1'-s_1+s_2'-s_2)$ and $y = s_1'-s_1-s_2'+s_2$, and we used the fact that $g_{\sigma}(s_1'-s_1) g_{\sigma}(s_2'-s_2) =  g_{\sigma}(\sqrt{2}x) g_{\sigma}(y/\sqrt{2})$.

 We separate Eq. (\ref{G2}) into a sum of two terms, one involving the pair of Feynman propagators and one involving the pair of Wightman functions. Integration over $x$ in the former term yields a proportionality factor $e^{-\sigma^2(E_1+E_2)^2}$; hence, this term is strongly suppressed. Integration over $y$ in the latter term yields a proportionality factor $\sqrt{16 \pi \sigma^2}\exp[-(E_2-E_1)^2 \sigma^2]$. Since $E_1 \sigma >> 1$ and $E_2 \sigma >> 1$, this factor is well approximated by $4 \pi \delta (E_1 - E_2)$. This implies that non-trivial correlations exist only for the detection of particles with the same energy. Eq. (\ref{G2}) then becomes
 \begin{eqnarray}
 G(E_1, \tau_1; E_2, \tau_2) = 4 \pi \alpha(E_1) \alpha(E_2) \delta (E_1-E_2) \int dS_1 dS_2 f_{\sigma}(\tau_1-S_1) f_{\sigma}(\tau_2 - S_2) \nonumber \\
\times  \int dx e^{-i(E_1+E_2)x} g_{\sigma}(\sqrt{2}x) \Delta^+ (x+S_2-S_1) \Delta^+ (x+S_1-S_2). \label{G3}
 \end{eqnarray}

\subsubsection{Uniformly accelerated detectors}
Next, we consider two uniformly accelerated detectors separated by constant proper distance $d$ in a direction normal to the direction of motion, that is, $x_1(\tau) = (\sinh(a \tau)/a,(\cosh(a \tau)-1)/a,0, 0 )$ and $x_2(\tau') =  (\sinh(a \tau'), (\cosh(a \tau')-1)/a, d, 0 )$. The detector's clocks are synchronized so that $x_1(0) = (0, 0, 0, 0)$ and $x_2(0) = (0, 0, d, 0)$.
 In this case,
 \begin{eqnarray}
 \Delta^+[x_1(\tau), x_2(\tau')] = - \frac{a^2}{16 \pi^2 \sinh[a(\tau - \tau'- i \tilde{\epsilon}-r)/2] \sinh[a(\tau - \tau'- i \tilde{\epsilon} + r)/2]}, \label{d12}
 \end{eqnarray}
where
\begin{eqnarray}
r = \frac{2}{a} \sinh^{-1}\left(\frac{ad}{2}\right)
\end{eqnarray}
is the proper time that it takes a light-ray emitted from a point of the path $x_1(\cdot)$ to reach a point of the path $x_2(\cdot)$.

We substitute Eq. (\ref{d12}) into Eq. (\ref{G3}), employ Eq. (\ref{eq1}) and integrate over $S_1 + S_2$, to obtain
\begin{eqnarray}
G(E_1, \tau_1; E_2, \tau_2) = \alpha(E_1) \alpha(E_2) \frac{a^4}{64 \pi^3} \delta (E_1 - E_2) \int dS f_{\sigma}(\Delta \tau - S) H(S), \label{G4}
\end{eqnarray}
where $S = S_2 - S_1$, $\Delta \tau = \tau_2 - \tau_1$, and
\begin{eqnarray}
H(S) = \int_{-\infty - i \tilde\epsilon}^{\infty - i \tilde\epsilon}  dx \frac{e^{-i(E_1+E_2)x} g_{\sigma}(\sqrt{2}x)}{\sinh[a(x-u)/2] \sinh[a(x+u)/2] \sinh[a(x+v)/2] \sinh[a(x-v)/2]}, \label{H}
\end{eqnarray}
In Eq. (\ref{H}), we wrote $u = S + r$ and $v = S - r$.

The function integrated in Eq. (\ref{H}) has 4 poles, at $x = \pm u$ and $x = \pm v$. The function $g_{\sigma}$ strongly suppresses the integral Eq. (\ref{H}), unless $|u|$ is of order $\sigma$ or smaller, or $|v|$ is of order $\sigma$ or smaller.

We  find analytic expressions that approximate $H(S)$ well in two regimes. The first regime corresponds to $|u - v| = 2r >> \sigma$. The distance between the two detectors is sufficiently large, so that the time delay of signal propagation between them is much larger than the temporal resolution $\sigma$.
 Then, the peaks at $|x| = u$ and $|x| = v$ do not overlap. Hence, the integral Eq. (\ref{H}) is well approximated by a sum of two terms, one corresponding to $|x| = u$ and one to $|x| = v$. To leading order in $1/(\sigma a)$,
\begin{eqnarray}
H(S) &=& \frac{g_{\sigma}(\sqrt{2}u)}{\sinh(aS) \sinh(ar)} \int_{-\infty - i \tilde\epsilon}^{\infty - i \tilde\epsilon} \frac{dx e^{-i(E_1+E_2)x}}{\sinh[a(x-u)/2] \sinh[a(x+u)/2]} \nonumber \\
&-& \frac{g_{\sigma}(\sqrt{2}v)}{\sinh(aS) \sinh(ar)} \int_{-\infty - i \tilde\epsilon}^{\infty - i \tilde\epsilon} \frac{dx e^{-i(E_1+E_2)x}}{\sinh[a(x-v)/2] \sinh[a(x+v)/2]}. \label{H2}
\end{eqnarray}
Carrying out the integrations in Eq. (\ref{H2}), we obtain
\begin{eqnarray}
H(S) = \frac{8\pi/a}{(e^{\frac{2\pi}{a}(E_1+E_2)}-1) \sinh(aS) \sinh(ar)} \left[ g_{\sigma}(\sqrt{2}(S+r)) \frac{\sin[(E_1+E_2)(S+r)]}{\sinh[a(S+r)]} \right. \nonumber \\
\left. - g_{\sigma}(\sqrt{2}(S-r)) \frac{\sin[(E_1+E_2)(S-r)]}{\sinh[a(S-r)]}\right]. \label{H3}
\end{eqnarray}
 We substitute Eq. (\ref{H3}) into Eq. (\ref{G4}) and integrate over $S$. In the Appendix, we evaluate this integral to leading order in $[(E_1+E_2)\sigma]^{-1}$ and $(a\sigma)^{-1}$. The result is
 \begin{eqnarray}
 G(E_1, \tau_1;E_2, \tau_2) = -\alpha(E_1)\alpha(E_2) \delta(E_1 - E_2) \frac{\frac{a^2}{8\pi}\tanh\left[\frac{\pi (E_1+E_2)}{2a}\right]}{(e^{\frac{2\pi}{a}(E_1+E_2)}-1)\sinh^2(a r)  }\nonumber \\
 \times \left[f_{\sigma}(\Delta \tau - r) + f_{\sigma}(\Delta \tau + r) \right]. \label{G5}
 \end{eqnarray}
The correlations are strongly peaked around $\Delta \tau = \pm r$, i.e., information about the recording of a particle in a detector is transmitted with the speed of light and affects the response of the other detector. Since Eq. (\ref{G5}) holds for $r >> \sigma$, it follows that $ar >> 1$, and thus,
\begin{eqnarray}
G(E_1, \tau_1;E_2, \tau_2) = -\frac{a^2}{2 \pi} \alpha(E_1)\alpha(E_2) \delta(E_1 - E_2) \frac{e^{-2ar}\tanh\left[\frac{\pi (E_1+E_2)}{2a}\right]}{e^{\frac{2\pi}{a}(E_1+E_2)}-1} \nonumber \\
\times \left[f_{\sigma}(\Delta \tau - r) + f_{\sigma}(\Delta \tau + r)\right]. \label{G6}
\end{eqnarray}
We observe that the correlations decrease exponentially with the time-delay $r$.

The other physically interesting regime for which an analytic expression for the correlations is obtained, corresponds to small separation between the detectors, $r << a^{-1}$. In this case, the effects of retarded propagation are insignificant, and the trajectories followed by the two detectors are indistinguishable. Hence, in this regime, the two detectors  may be viewed as   independent subcomponents of a single macroscopic detector. At the limit
  $r \rightarrow 0$, Eq. (\ref{H}) becomes
\begin{eqnarray}
H(S) = \int_{-\infty - i \tilde{\epsilon}}^{\infty - i \tilde{\epsilon}} \frac{dx e^{-i(E_1+E_2)x} g_{\sigma}(\sqrt{2}x)}{\sinh^2[a(x-S)/2]\sinh^2[a(x+S)/2]}. \label{H4}
\end{eqnarray}
The presence of the function $g_{\sigma}$ implies that the integral is strongly suppressed, unless $S$ is of order $\sigma$ or smaller. To leading order in $1/(\sigma a)$, we obtain
\begin{eqnarray}
H(S) &=& \frac{16 \pi g_{\sigma}(\sqrt{2}S)}{a^2(e^{\frac{2\pi}{a}(E_1+E_2)}-1)\sinh^2(aS)}
\nonumber \\
&\times& \left[(E_1+E_2) \cos[(E_1+E_2)S]-a \coth(aS) \sin[(E_1+E_2)S]\right]. \label{H5}
\end{eqnarray}
We substitute Eq. (\ref{H5}) into Eq. (\ref{G4}), and integrate over $S$. To leading order in $[(E_1+E_2)\sigma]^{-1}$ and $(a\sigma)^{-1}$, we obtain (see the Appendix)
\begin{eqnarray}
G(E_1, \tau_1;E_2, \tau_2) = - \frac{(E_1+E_2)^2}{ 8 \pi} \alpha(E_1) \alpha(E_2) \delta(E_1-E_2) \frac{\coth\left[\frac{(E_1+E_2)\pi}{2a}\right]}{e^{\frac{2\pi}{a}(E_1+E_2)}-1} f_{\sigma}(\Delta \tau). \label{G7}
\end{eqnarray}

\paragraph{Second-order coherence functions.}
Eqs. (\ref{G6}) and (\ref{G7}) provide the two-time correlations for particle detection for a pair of uniformly accelerated detectors in the regimes $r \sigma >> 1$ and $r a << 1$ respectively. It is convenient to express these correlations in terms of the second-order coherence function $g^{(2)}$, which is standardly employed in quantum optics. For a stationary field configuration, the second-order coherence function is defined
 as
\begin{eqnarray}
g^{(2)}(\Delta \tau) = \frac{\langle I(0)I(\Delta \tau)\rangle}{I(0)I(\Delta \tau)}. \label{2ndcor}
\end{eqnarray}

If $g^{(2)}(\Delta \tau) < g^{(2)}(0)$, there is a higher probability for a simultaneous detection of a pair of particle, whence the particles are said to be {\em bunching}. If  $g^{(2)}(\Delta \tau) > g^{(2)}(0)$, the particles are said to be anti-bunching.
 When considering separated detectors, we have to take into account the effect of retarded propagation, so the natural definition of bunching is $g^{(2)}(\Delta \tau) < g^{(2)}(r)$ for $|\Delta \tau |> r$.

Also, the statistics of particle detection are sub-Poissonian if  $ g^{(2)}(0) < 1$, Poissonian if $ g^{(2)}(0) = 1$, and super-Poissonian if  $ g^{(2)}(0) < 1$.

The intensity $I$ and the intensity correlations are defined  by  Eq. (\ref{I}) and Eq. (\ref{II}), respectively. Using Eqs. (\ref{pg}), (\ref{G6}) and (\ref{G7}), we obtain
\begin{eqnarray}
g^{(2)}(\Delta \tau) &=& 1 - 2 \pi a^2 e^{-2ar} \frac{\int dE E^2 \alpha(E)^2 e^{-\frac{2\pi E}{a}}\mbox{sech}^2(\pi E/a)}{\left[\int dE \alpha(E) E^2/(e^{\frac{2 \pi E}{a}}-1)  \right]^2}
\nonumber \\
&\times& \left[f_{\sigma}(\Delta \tau - r) + f_{\sigma}(\Delta \tau + r)\right], \mbox{for} \;\; r >> \sigma \hspace{1cm} \label{2ndcora}\\
g^{(2)}(\Delta \tau) &=& 1 - \frac{\pi \int dE E^4 \alpha(E)^2e^{-\frac{2\pi E}{a}}\mbox{cosech}^2(\pi E/a) }{2 \left[\int dE \alpha(E) E^2/(e^{\frac{2 \pi E}{a}}-1)  \right]^2} f_{\sigma}(\Delta \tau), \mbox{for} \;\;r = 0. \label{2ndorb}
\end{eqnarray}

 An important special case is that of a macroscopic two-level detector, namely, a  detector recording only a narrow frequency band $[E-\frac{1}{2}\Delta E, E + \frac{1}{2}\Delta E]$, with $\sigma^{-1} << \Delta E << E$. In this case,
\begin{eqnarray}
g^{(2)}(\Delta \tau)  &=& 1 - \frac{8\pi a^2}{E^2 \Delta E} e^{-2ar} \tanh(\pi E/a) \left[f_{\sigma}(\Delta \tau - r) + f_{\sigma}(\Delta \tau + r)\right],  \mbox{for} \;\; r >> \sigma \\
g^{(2)}(\Delta \tau) &=& 1 - \frac{2 \pi}{\Delta E} f_{\sigma}(\Delta \tau), \mbox{for} \;\; r =0.
\end{eqnarray}

It is evident from the equations above that particles recorded by a uniformly accelerated detector anti-bunch and satisfy sub-Poissonian statistics, at both regimes $r \sigma >> 1$ and $r \rightarrow 0$.

\subsubsection{Static detectors in a thermal bath}

In order to examine whether the relation between uniform acceleration and temperature holds, we must construct the second-order coherence function corresponding to a pair of static Unruh-DeWitt detectors when the field is in a thermal state of temperature $T$. To this end, we compare the positive frequency Wightman function for a thermal state Eq. (\ref{deltab}) with the one corresponding to a uniformly accelerated observer Eq. (\ref{d12}). One obvious difference is the presence of the delta functions in Eq. (\ref{deltab}); these, however, do not contribute to Eq. (\ref{G3}) because they vanish along the real axis. We denote the regular part of the thermal Wightman function  Eq. (\ref{deltab}) as $\tilde{\Delta}_{\beta}$ and the Wightman function for the accelerated observers Eq. (\ref{d12}), evaluated for $a = 2\pi/\beta$  as $\Delta_{a = 2\pi/\beta}$. Then,
\begin{eqnarray}
\tilde{\Delta}_{\beta} = \frac{\beta}{2 \pi r} \sinh\left(\frac{2 \pi r}{\beta}\right)\Delta_{a = 2\pi/\beta}. \label{comparison}
\end{eqnarray}

In Eq. (\ref{comparison}), $r$ refers to the time delay due to retarded propagation. For static detectors in a thermal bath, $r$ coincides with their spatial separation. Thus, the difference in the two Wightman's function is a multiplicative factor $e(r) = \frac{\beta}{2 \pi r} \sinh\left(\frac{2 \pi r}{\beta}\right)$ that tends to unity at the limit $r \rightarrow 0$. This implies that the function $G(E_1, \tau_1; E_2, \tau_2)$ for static  detectors in a thermal field is obtained from Eqs. (\ref{G6}) and (\ref{G7}) by setting $a = 2\pi/\beta$ and multiplying by a factor $e(r)^2$.

In the regime  $r >> \sigma$,
\begin{eqnarray}
G(E_1, \tau_1; E_2, \tau_2) &=& - \frac{1}{8 \pi r^2} \alpha(E_1) \alpha(E_2)  \delta (E_1 - E_2) \nonumber \\
&\times&
\frac{\tanh[\beta(E_1+E_2)/4]}{e^{\beta(E_1+E_2)}-1} \left[f_{\sigma}(\Delta \tau - r) + f_{\sigma}(\Delta \tau + r)\right], \label{GT2}
\end{eqnarray}
and for  $r = 0$,
\begin{eqnarray}
G(E_1, \tau_1;E_2, \tau_2) = - \frac{(E_1+E_2)^2}{8\pi} \alpha(E_1) \alpha(E_2) \delta(E_1-E_2) \frac{\coth[\beta(E_1+E_2)/4]}{e^{\beta(E_1+E_2)}-1} f_{\sigma}(\Delta \tau).
\end{eqnarray}

Then, the second-order coherence functions are
\begin{eqnarray}
g^{(2)}(\Delta \tau) &=& 1 - \frac{\pi}{2 r^2} \frac{\int dE E^2\alpha(E)^2 e^{-\beta E} \mbox{sech}^2(\beta E/2) }{[\int dE E^2 \alpha(E)/(e^{\beta E} - 1)]^2}
\nonumber \\
&\times&
\left[f_{\sigma}(\Delta \tau - r) + f_{\sigma}(\Delta \tau + r)\right], \mbox{for} \; r >> \sigma \hspace{1cm} \\
g^{(2)}(\Delta \tau) &=& 1- \frac{\pi \int dE E^4\alpha(E)^2 e^{-\beta E} \mbox{cosech}^2(\beta E/2)}{2[\int dE E^2 \alpha(E)/(e^{\beta E} - 1)]^2}f_{\sigma}(\Delta \tau), \mbox{for}\;r = 0. \label{g2Tb}
\end{eqnarray}

For a macroscopic two-level detector,
\begin{eqnarray}
g^{(2)}(\Delta \tau) &=& 1 - \frac{2 \pi}{r^2E^2\Delta E} \tanh(\beta E/2) \left[f_{\sigma}(\Delta \tau - r) + f_{\sigma}(\Delta \tau + r)\right], \mbox{for} \; r >> \sigma \hspace{1cm}\\
g^{(2)}(\Delta \tau) &=& 1 - \frac{2 \pi}{\Delta E}f_{\sigma}(\Delta \tau), \mbox{for}\;r = 0
\end{eqnarray}

Comparing Eq. (\ref{g2Tb}) and Eq. (\ref{2ndorb}) we see that for $r = 0$, the intensity fluctuations recorded by  uniformly accelerated detectors and the ones recorded by  static detectors in a thermal bath are fully equivalent, given  Unruh's relation between temperature and acceleration $T = \frac{a}{2\pi}$.
  However, this equivalence is lost for separated detectors, in the regime $r >> \sigma$.  While in both cases we have particle anti-bunching, the dependence of the correlations on the time delay $r$ differs significantly. A pair of inertial detectors in a thermal bath has long range correlations (that fall-off with $r^{-2}$), while a pair of accelerated observers has short-range correlations that drop exponentially with r.

  It is important to emphasize that the difference above arises at the level of the field's two-point functions, which are expected to enter into the correlations for any observable defined with respect to the two separated detectors. One therefore expects that the correlations of separated accelerated observers will {\em generically} deviate from the thermal ones.

\subsection{Relation to the Glauber detector}
The fact that the second-order coherence function (\ref{g2Tb}) for the thermal state exhibits anti-bunching ($g^{(2)}(\Delta \tau) > g^{(2)}(0)$ for $r=0$) might be viewed as contradictory to the well-known fact that thermal photons bunch \cite{WM}. The difference here is that we consider  Unruh-DeWitt detectors, while the bunching behavior of photons in quantum optics is derived for Glauber detectors.

In Glauber-type detectors, the $n$-time detection intensity correlation function at different spacetime points $x_1, x_2, \ldots, x_n$ is proportional to
 \begin{eqnarray}
G^{(n)}(x_1, x_2, \ldots, x_n) = Tr \left(\hat{\phi}^{(-)}(x_1)  \ldots\hat{\phi}^{(-)}(x_n) \hat{\phi}^{(+)}(x_n) \ldots \hat{\phi}^{(-)}(x_1) \hat{\rho}_0 \right), \label{gn}
\end{eqnarray}
where $\hat{\phi}^{(-)}$ and $\hat{\phi}^{(+)}$ are respectively the positive- and negative frequency components of the Heisenberg-picture scalar field $\hat{\phi}(x)$, and $\hat{\rho}_0$ is the field's density matrix.

The correlation functions of the Glauber detectors can be obtained from the general theory of Sec. 2, in particular, Eq. (\ref{ptn}). The interaction Hamiltonian for a Glauber detector is of the form
\begin{eqnarray}
\hat{H}_I = \sigma^+ \hat{\phi}^{(-)}(x) + \hat{\sigma}^- \hat{\phi}^{(+)}(x),
\end{eqnarray}
where $\hat{\sigma}^-\hat{P}_0 = \hat{P}_0 \hat{\sigma}^- = 0$.

It is possible to bring the Unruh-Dewitt interaction Hamiltonian into the Glauber form through a redefinition of variables and the consideration of `dressed' states for field and detector \cite{CoPi}. For this reason, we believe that the interaction Hamiltonian does not constitute the most important difference between the two types of detector. As explained below, the most important difference is the scale of the temporal coarse-graining parameter $\sigma$.

The two-time probability distribution Eq. (\ref{pd12}) for a pair of  detectors then becomes
\begin{eqnarray}
p(E_1,\tau_1;E_2, \tau_2) = \alpha(E_1)\alpha(E_2)\int ds_1 ds_1' ds_2 ds_2' \times e^{-iE_1(s_1'-s_1) - iE_2(s_2'-s_2)}
 \nonumber \\
\times \sqrt{f_{\sigma_1}(t_1-s_1)f_{\sigma_1}(t_1-s_1') f_{\sigma_2}(t_2-s_2) f_{\sigma_2}(t_2 - s_2')} \hspace{2.5cm}  \nonumber \\
\times  Tr_{{\cal H}_q} \left[ \hat{\rho}_0 {\cal A} \left( \hat{\phi}^{(-)}[x_1(s_1)] \hat{\phi}^{(-)}[x_2(s_2')]\right) {\cal T} \left(\hat{\phi}^{(+)}[x_2(s_2)] \hat{\phi}^{(+)}[x_1(s_1)]\right) \right], \label{pe22}
\end{eqnarray}
where $x_1(\cdot), x_2(\cdot)$ are the detector's world-lines.

Eq. (\ref{pe22}) leads to a two-time probability proportional to $G^{(n)}$ of Eq. (\ref{gn}) at the limit $\sigma \rightarrow 0$. This means that the resolution timescale $\sigma$ of the detector must be smaller than any time-scale appearing in the correlation function
\begin{eqnarray}
Tr_{{\cal H}_q} \left[ \hat{\rho}_0 {\cal A} \left( \hat{\phi}^{(-)}[x_1(s_1)] \hat{\phi}^{(-)}[x_2(s_2')]\right) {\cal T} \left(\hat{\phi}^{(+)}[x_2(s_2)] \hat{\phi}^{(+)}[x_1(s_1)]\right) \right].
 \end{eqnarray}
 In contrast, an Unruh-Dewitt detector recovers the Planckian spectrum only if
 $\sigma a >> 1$, i.e., if the resolution timescale is much larger than the characteristic timescales of the correlation function.

\section{Conclusions}

In this paper, we  constructed new models for macroscopic particle detectors, where the time of transition is treated as a quantum observable. These models allow for a precise treatment of temporal fluctuations and correlations in particle detection. For  Unruh-DeWitt detectors, moving along generic trajectories in Minkowski spacetime and coupled to a quantum field,  we obtained a causal, and approximately local-in-time expression for the detector's response:  the response function depends only on properties of the path that are defined at time $t$, within an accuracy allowed by the detector's resolution. This implies that Unruh's relation between acceleration and temperature is quite robust and extends beyond the case of uniform acceleration: slow changes of the detector's acceleration result to corresponding changes in the temperature of the recorded Planckian spectrum.

 Our most important result is the explicit calculation of the two-time probabilities and of the intensity fluctuations for uniformly accelerated detectors. We find that that intensity fluctuations recorded by a single detector are thermal: the second-order coherence function of a uniformly accelerated Unruh-DeWitt detector is identical to that of a static detector in a thermal bath.  This implies that the records of particle detection by a single Unruh-Dewitt detector cannot distinguish
  whether the detector is accelerated, or whether it is static in the presence of
 a thermal bath.
In contrast, the intensity correlations for  a pair of Unruh-DeWitt detectors, with the same linear acceleration, but moving along different trajectories are non-thermal. Hence, records of particle detection from a pair of detectors, at rest with respect to each other, do distinguish whether the detectors are in an accelerated reference frame or whether they are static and in contact with the thermal bath. This result greatly strengthens the view of the Unruh effect as a fundamentally local phenomenon \cite{FU}.


It is important to note that the existence of non-thermal correlations cannot be inferred from the quantum-field theoretic treatment of the Unruh effect (Fulling-Rindler quantization). In that treatment,
the restriction of the Minkowski vacuum in one Rindler wedge is a thermal state with respect to the Rindler time coordinate that is associated to accelerated observers. If this restricted state were interpreted as the quantum state of the field in the accelerated reference frame, then one would expect that all field observables with support on
one Rindler wedge  would exhibit thermal properties in the accelerated frame, irrespective of  whether these observables are local or not. Our results suggest
that the  Fulling-Rindler quantization should not be taken as a literal construction of the quantum state in an accelerated reference frame \cite{BKNM}. Indeed, the interpretation of the Fulling-Rindler vacuum has no bearing on the, fundamentally local, physics of the Unruh effect \cite{FU}. In contrast, the detectors employed here define
physically meaningful observables for the quantum field, with a natural interpretation in terms of accelerated frames of reference.   Hence, the conclusion that the particle-detection correlations  are not thermal   provides novel information that cannot be extracted from the procedure of Fulling-Rindler quantization, at least in a straightforward way.

The question then arises how one should describe the quantum field state in the reference frame of accelerated observers. In our opinion, the most promising approach would involve the framework of local quantum physics \cite{haag}, with a construction of an algebra of local observables and the associated space of states. In order to explain this point, we note that the $n$-time joint probabilities Eq. (\ref{ptn}) can be expressed in terms of positive operators $\hat{P}_n(E_1,t_1, E_2, t_2, \ldots E_n, t_n)$  on the  Hilbert space of the quantum field, by tracing out the detectors' degrees of freedom. When considering detectors moving along the same trajectory, the operators $\hat{P}_n(E_1,t_1, E_2, t_2, \ldots E_n, t_n)$ define a subset of local observables with thermal expectation values. This description fits naturally with the idea of locally thermal states, defined in Ref. \cite{buch}. In this work, states of local thermal equilibrium are defined if the expectation value of a   sufficiently large number of thermal observables in the vicinity of a spacetime point $x$ coincide with the expectation value of a globally KMS state at this point. This analogy suggests the possibility of describing the quantum field state in an accelerating reference frame as a local equilibrium state, where the local temperature at each point is proportional to the proper acceleration in this frame, but with non-local correlations between different spacetime points.

 The fact that there may exist significant information about the correlation functions of the field that is not captured by transformations between different field vacua is an important observation that could be relevant to the information-loss paradox in black holes.   There are, of course, significant differences between the physical set-ups of the Unruh and the Hawking effects, so the results of the present paper cannot be directly applied to  black holes. Nonetheless, the point of principle is intriguing. The information-loss paradox arises from the assumption that the final state of the field at late times is globally Gibbsian, hence, mixed. If, however, the state is non Gibbsian, but only locally thermal, it is at least conceivable that the information of the initial state may persist throughout the process of black hole formation and evaporation, in the form of non-local correlations of the field variables.

\begin{appendix}
\section{Evaluation of integrals}

Here, we provide the derivation of Eqs. (\ref{G5}) and (\ref{G7}) from the evaluation of integrals of the form   Eq. (\ref{G4}).

In order to derive Eq. (\ref{G5}), we must compute the integral
\begin{eqnarray}
\int dS f_{\sigma}(S - \Delta \tau) H(S),
\end{eqnarray}
where $H(S)$ is given by Eq. (\ref{H3}). This means that we must evaluate integrals of the form
\begin{eqnarray}
J_1 = \int \frac{dS}{\sinh(aS)} f_{\sigma}(S - \Delta \tau) g_{\sigma}(\sqrt{2}(S \pm r)) \frac{\sin[E(S \pm r)]}{\sinh[a(S \pm r)]}
\end{eqnarray}
For concreteness we   compute the integral above with the (-) sign.
The integrand is peaked around $S = \Delta  \tau $ and $S =  r$.  We change the integration variable to $x = S- r$, so that
\begin{eqnarray}
J_1 =  \int \frac{dx}{\sinh[a(x+r)]} f_{\sigma}(x - v) g_{\sigma}(\sqrt{2}x)\frac{\sin(Ex)}{\sinh(ax)}, \label{j1a}
\end{eqnarray}
where $v = \Delta \tau - r$.
We set $A(x) = f_{\sigma}(x - v) g_{\sigma}(\sqrt{2}x)/\sinh[a(x+r)]$ and $B(x) = \frac{\sin(Ex)}{\sinh(ax)}$. Denoting by $\tilde{A}(k)$ and $\tilde{B}(k)$ the Fourier transforms of $A(x)$ and $B(x)$ respectively, Eq. (\ref{j1a}) becomes
\begin{eqnarray}
J_1 =  \int \frac{dk }{2\pi} \tilde{A}(k) \tilde{B}^*(k). \label{j1b}
\end{eqnarray}
 For $r >> \sigma$, the $\sinh[a(x+r)]$ term is well approximated by $\sinh(ar)$, and, consequently,
  $\tilde{A}(k)$   is proportional to $e^{-\sigma^2 k^2/3}$. This means that values of $k >> \sigma^{-1}$ are suppressed in the integration Eq. (\ref{j1b}). The function $B(x)$ is oscillatory with a frequency of order $E$; hence, to leading order in $(E\sigma)^{-1}$, we can substitute $\tilde{B}^*(k)$ with $\tilde{B}(0)$ in Eq. (\ref{j1b}). Thus,
\begin{eqnarray}
J_1 \simeq \tilde{B}(0) \int \frac{dk }{2\pi} \tilde{A}(k) = \tilde{B}(0)A(0).
\end{eqnarray}
We compute
\begin{eqnarray}
\tilde{B}(0) = \int dx \frac{\sin(Ex)}{\sinh(ax)} = \frac{\pi}{a} \tanh\left[\frac{\pi E}{2a}\right].
\end{eqnarray}
Hence,
\begin{eqnarray}
J_1 = \frac{\pi}{a \sinh(a r)} \tanh\left[\frac{\pi E}{2a}\right] f_{\sigma}(v),
\end{eqnarray}
from which Eq. (\ref{G5}) follows.

The derivation of Eq. (\ref{G7}) proceeds along similar lines. It involves an integral of the form
\begin{eqnarray}
J_2 = \int \frac{dS}{\sinh^2(aS)} f_{\sigma}(S - \Delta \tau) g_{\sigma}(\sqrt{2}S) \left[E \cos(ES)-a \coth(aS) \sin(ES)\right]
\end{eqnarray}
Using the same arguments as before, we find that to leading order in $(E\sigma)^{-1}$,
$J_2 =   f_{\sigma}(\Delta \tau) \tilde{B}(0)$,
where now
\begin{eqnarray}
\tilde{B}(0) = \int dS \frac{E \cos(ES)-a \coth(aS) \sin(ES)}{\sinh^2(aS)} = - \frac{\pi E^2 \coth\left(\frac{\pi E}{2a}\right)}{2 a^2}.
\end{eqnarray}
Hence,
\begin{eqnarray}
J_2 = - \frac{\pi E^2 \coth\left(\frac{\pi E}{2a}\right)}{2a^2} f_{\sigma}(\Delta \tau),
\end{eqnarray}
from which Eq. (\ref{G7}) follows.
\end{appendix}

\end{document}